\documentclass[twocolumn]{aastex631}
\usepackage{amsmath}

\shorttitle{Noise2Astro}
\shortauthors{Zhang et al.}

\graphicspath{{./}{figures/}}

\begin{document}

\title{Noise2Astro: Astronomical Image Denoising With Self-Supervised Neural
Networks}

\correspondingauthor{Yunchong Zhang}
\email{aruba19th@uchicago.edu}

\author[0000-0001-6454-1699]{Yunchong Zhang}
\affiliation{ Department of Astronomy and Astrophysics, University of Chicago, 5640 South Ellis Avenue, Chicago, IL 60637, USA}
\author[0000-0001-6706-8972]{Brian Nord}
\affiliation{ Department of Astronomy and Astrophysics, University of Chicago, 5640 South Ellis Avenue, Chicago, IL 60637, USA}
\affiliation{ Fermi National Accelerator Laboratory, P.O. Box 500, Batavia, IL 60510, USA}
\affiliation{ Kavli Institute for Cosmological Physics, University of
Chicago, 5640 South Ellis Avenue, Chicago, IL 60637, USA}
\author[0000-0002-6015-8614]{Amanda Pagul}
\affiliation{Department of Physics and Astronomy, University of California Riverside, Pierce Hall, Riverside, CA, 92521 USA}
\author[0000-0003-1135-6419]{Michael Lepori}
\affiliation{Department of Computer Science, Brown University, 115 Waterman St, Providence, RI 02906, USA}

\begin{abstract}

In observational astronomy, noise  obscures signals of interest. Large-scale astronomical surveys are growing in size and complexity, which will produce more data and increase the workload of data processing. Developing automated tools, such as convolutional neural networks (CNN), for denoising has become a promising area of research. We investigate the feasibility of CNN-based self-supervised learning algorithms (e.g., Noise2Noise) for denoising astronomical images. We experimented with Noise2Noise on simulated noisy astronomical data. We evaluate the results based on the accuracy of recovering flux and morphology. This algorithm can well recover the flux for Poisson noise ( $98.13${\raisebox{0.5ex}{\tiny$^{+0.77}_{-0.90} $}$\large\%$}) and for   Gaussian noise when image data has a smooth signal profile ($96.45${\raisebox{0.5ex}{\tiny$^{+0.80}_{-0.96} $}$\large\%$}).

\end{abstract}

\keywords{Neural networks (1933) --- Astronomy data reduction (1861) --- Astronomy image processing (2306)}

\section{Introduction} \label{sec:intro}

Deep learning methods for image denoising are dependent on the ground-truth data provided in the training set, and most efforts have been directed toward refining neural network (NN) architectures \citep[e.g.,][]{astrounet,radiosource}. To generate training data, we either need a thorough and robust understanding of our astronomical objects to simulate the ground-truth signal or real data with high signal-to-noise garnered from long exposures in observations. 

\begin{figure*}[tbph!] \label{fig1}
\includegraphics[width = \linewidth]{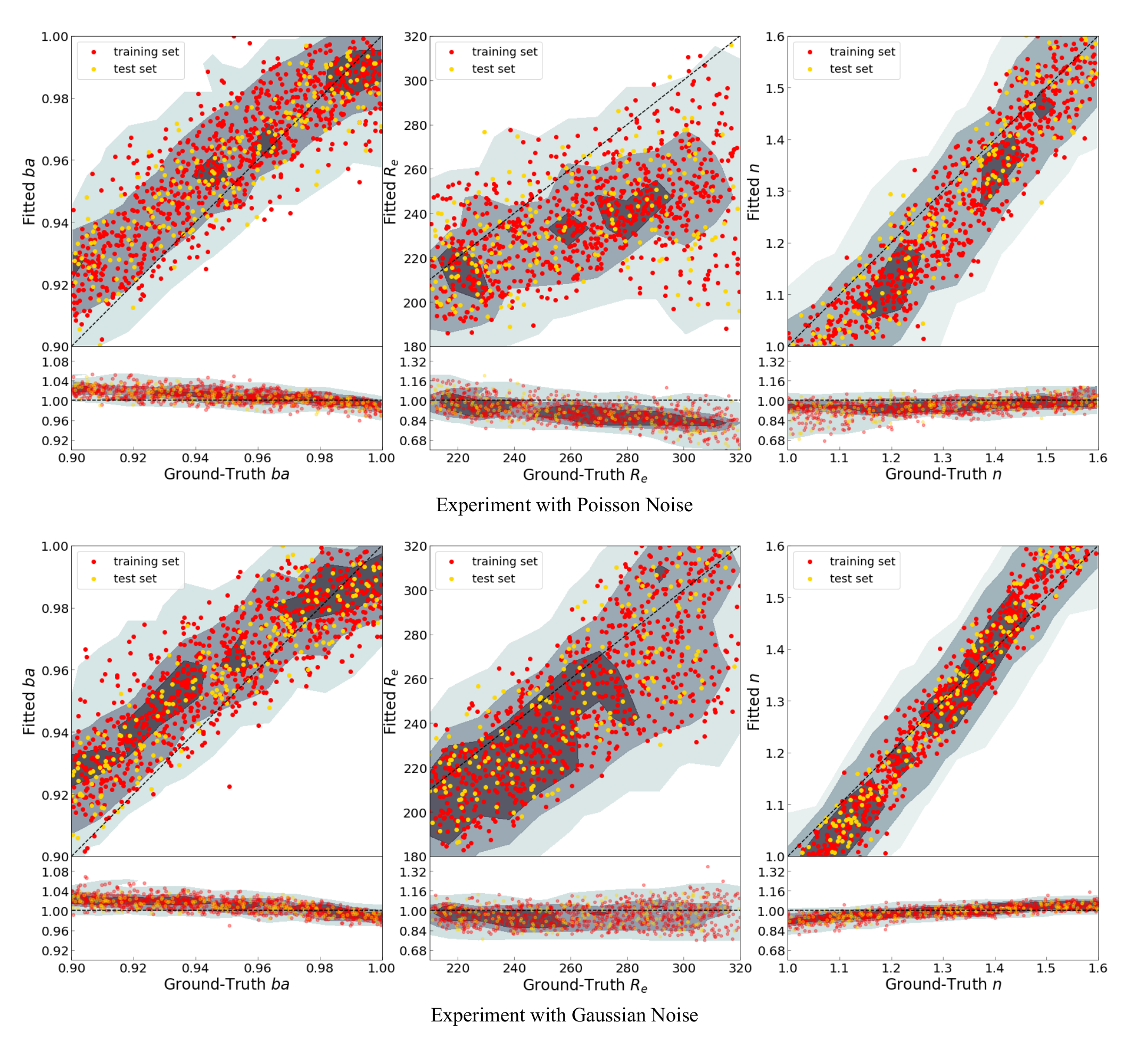}
          
\caption{Upper panels in top and bottom row: morphological parameters comparison between fitted results and ground truths; Lower panels in top and bottom row: the distribution of morphological parameters $Parameter_{fit}/Parameter_{true}$. The filled contours reflect the bin counts of instances. The darker the color, the denser the distribution of instances.}
\end{figure*}

\cite{Noise2Noise} demonstrate that denoising images with only noisy images -- a self-supervised process -- is possible. Several works \citep{Noise2Self,Noise2Void} have explored the feasibility of blind denoising using self-supervision. Astronomical imaging often generates multiple exposures of the same objects, which can be used with Noise2Noise. By exploiting the self-similarity in natural images, the Noise2Noise algorithm introduces a general method for finding an optimal denoiser for a given dataset. For a given dataset with noisy image pairs, suppose each image pair contains two noisy images $x_1$ and $x_2$ of the same dimensions, such that $x_1 = y + n_1$ and $x_2 = y + n_2$, where $y$ is the ground truth, and $n_1$ and $n_2$ are two different additive noise components. Assuming each noise component $n_i$ is independent from the ground truth, one can find an optimal transformation $f$ from $x_1$ to $x_2$ among an entire class of transformations $f_{\theta}$, where the transformation is parameterized by $\theta$, by minimizing the self-supervised loss between the two noisy images: 
\begin{equation}
\mathcal{L} = \mathbb{E}||f_{\theta}(x_1) - x_2||. 
\end{equation}
Due to the independence of the noise components, one expects the transformation $f$ to fail in predicting the noisy component $n_2$ in $x_2$ when only given $x_1$ as input. Therefore, the optimal transformation $f$ would only predict the ground truth component $y$ and become an effective denoiser to images within the distribution of the training data.

\section{Method} \label{sec:method}
For Experiment 1, we applied the algorithm to two architectures: U-net \citep{U-net} and a denoising convolutional neural network (DnCNN) \citep{DNCNN}. 

We simulate training and test data. The signal components of each simulated image pair are generated as Sérsic profiles \citep{Sersic_1963} using the parametric fitting tool GALFIT \citep{Peng_2010} by inputting the chosen Sérsic index $n$, half-light radius $R_e$, positional angle $PA$, semi-major-axis to semi-minor-axis ratio $ba$, and magnitude $M$. We randomly sample from a range for each parameter except for the magnitude, which we randomly sample from the DES public catalog \citep{DES} processed by the Weak-lensing-Deblending software \citep{weaklensingdeblend, wld_software}. We simulate the noise components using GALSIM \citep{Galsim} and combine them with signal components for the training set. 

To train networks with self-supervision, we generate a pair of images with the same signal but different noise as one instance. For Experiment 1, we generate 1000 $128 \rm{pix} \times 128 \rm{pix}$ instances with Poisson noise, and we choose relatively wide ranges of morphological parameters ($n \in [0.5,6]; ba \in [0.01,1]; R_e \in [4,20]$) for the signal. For Experiment 2, we generate 1000 $256 \rm{pix} \times  256\rm{pix}$ image instances with narrower ranges of morphological parameters ($n \in [1.0,1.6]; ba \in [0.9,1]; R_e \in [200,320]$) to simulate low-surface brightness signal components. We generate two train-test sets for Experiment 2: one has entirely Gaussian noise components, and the other has Poisson noise and the same signal components as the former one. The Gaussian noise is simulated with a sigma taken from the value of mean sky-level per pixel from the DES public catalog \citep{DES}. For all experiments, we use the first 800 instances for training and all 1000 instances for testing.

\section{Results} \label{sec:results}
We evaluate Experiment 1 by measuring the bias image: $I_{bias} = \frac{I_{output}}{I_{true}}$. Out of the two network architectures, the U-net produces the least bias. While all  instances in Experiment 1 tend to have high biases in the image region where the signal tends to be peaky (relatively high surface brightness), the network output in the image region where the signal is smooth (relatively low surface brightness) converges to the ground truth.

In Experiment 2, we implement the U-net and examine the recovery of flux and signal morphology. To assess the morphological accuracy, we performed 2D Sérsic function fitting on our outputs with SciPy \citep{2020SciPy-NMeth}. We mask out pixels with a relatively large bias for fitting by choosing a circular mask with radius of $5 \rm{pix}$ centered at the peak of the signal. We also applied constraints to the $x$ and $y$ positions of the 2D Sérsic function, while other parameters are allowed to vary. For instances with Poisson (Gaussian) noise, the network recovers $98.13${\raisebox{0.5ex}{\tiny$^{+0.77}_{-0.90} $}$\large\%$} ($96.45${\raisebox{0.5ex}{\tiny$^{+0.80}_{-0.96} $}$\large\%$}) of the flux.

We compare outputs from our analysis (network denoising and Sérsic fitting) with the ground truth in Figure \ref{fig1}. In the lower panels, we show the bias ($P_{bias} = Parameter_{fit}/Parameter_{true}$) versus ground truth for each parameter. For each parameter, if the biases of experiment instances have a narrower distribution, the fitted values from the overall routine are in better agreement with the ground truths. The fitted $ba$ agrees with ground truths in both Gaussian and Poisson noise experiments ($ba_{ bias} \simeq 0.96 \sim 1.04$). For $n$, fitted values agree better with ground truths when the noise is Gaussian ($n_{ bias} \simeq 0.86 \sim 1.1$) than when the noise is Poisson ($n_{ bias} \simeq 0.7 \sim 1.14$). For $R_e$, the fitted values poorly agree with ground truths in both Gaussian ($R_{e\ bias} \simeq 0.72 \sim 1.2$) and Poisson cases ($R_{e\ bias} \simeq 0.6 \sim 1.16$), and few instances are on the line where $P_{bias} = 1$. In the case of Poisson noise, the $R_{e}$ tends to be significantly underestimated ($median(R_{e\  bias}) \simeq 0.84$). Considering that the ground-truth $R_e$ is much larger than the size of images in the test set, such a result can be expected.

We distinguish between the distribution of training-set and test-set instances in Figure \ref{fig1}, where ``test-set" refers to the 200 instances of the total 1000 testing instances that are not exposed to the model during training. In the two experiments, we find no obvious deviation of the ``test-set" instance distribution from that of the training-set. 
 
We conclude that NNs with self-supervision can achieve photometric accuracy within $5\%$ of the ground truth in limited scenarios: the self-training set has to contain images with sufficient resemblance and the objects should have light profiles with low flux gradients. Therefore, this algorithm would potentially be applicable to denoise faint, diffuse objects.

\section*{Acknowledgements}

We acknowledge the Deep Skies Lab as a community of multi-domain experts and collaborators who’ve facilitated an environment of open discussion, idea-generation, and collaboration. This community was important for the development of this project.

This manuscript has been authored by Fermi Research Alliance, LLC under Contract No. DE-AC02-07CH11359 with the U.S. Department of Energy, Office of Science, Office of High Energy Physics.

\bibliography{sample631}{}
\bibliographystyle{aasjournal}

\end{document}